\pdfoutput=1
\RequirePackage{ifpdf}
\ifpdf 
\documentclass[pdftex]{sigma}
\else
\documentclass{sigma}
\fi

\def\p{\partial}

\numberwithin{equation}{section}
\newtheorem{Theorem}{Theorem}[section]
\newtheorem{Lemma}[Theorem]{Lemma}
{ \theoremstyle{definition}
\newtheorem{Example}[Theorem]{Example}
}

\begin{document}

\allowdisplaybreaks

\renewcommand{\thefootnote}{$\star$}

\renewcommand{\PaperNumber}{006}

\FirstPageHeading

\ShortArticleName{The Master $T$-Operator for Inhomogeneous $XXX$ Spin Chain and mKP Hierarchy}

\ArticleName{The Master $\boldsymbol{T}$-Operator for
Inhomogeneous\\ $\boldsymbol{XXX}$ Spin Chain and
mKP Hierarchy\footnote{This paper is a~contribution to the Special Issue in honor of
Anatol Kirillov and Tetsuji Miwa.
The full collection is available at
\href{http://www.emis.de/journals/SIGMA/InfiniteAnalysis2013.html}{http://www.emis.de/journals/SIGMA/InfiniteAnalysis2013.html}}}

\Author{Anton ZABRODIN~$^{\dag^1\dag^2\dag^3\dag^4}$}

\AuthorNameForHeading{A.~Zabrodin}

\Address{$^{\dag^1}$~Institute of Biochemical Physics,
4 Kosygina, 119334, Moscow, Russia}

\Address{$^{\dag^2}$~ITEP, 25 B.~Cheremushkinskaya, 117218, Moscow, Russia}
\EmailDD{\href{mailto:zabrodin@itep.ru}{zabrodin@itep.ru}}

\Address{$^{\dag^3}$~National Research University Higher School of Economics,\\
\hphantom{$^{\dag^3}$}~20 Myasnitskaya Ulitsa, Moscow 101000, Russia}

\Address{$^{\dag^4}$~MIPT, Institutskii per.~9, 141700, Dolgoprudny, Moscow region, Russia}

\ArticleDates{Received October 18, 2013, in f\/inal form January 08, 2014; Published online January 11, 2014}

\Abstract{Following the approach of
[Alexandrov A., Kazakov V., Leurent S., Tsuboi Z., Zabrodin A., \textit{J.~High Energy Phys.}
  \textbf{2013} (2013), no.~9, 064, 65~pages, arXiv:1112.3310], we show how to construct the master $T$-operator
for the quantum
inhomogeneous ${\rm GL}(N)$ $XXX$ spin chain with twisted
boundary conditions.
It satisf\/ies
the bilinear identity and Hirota equations
for the classical mKP hierarchy.
We also characterize the class
of solutions to the mKP hierarchy
that correspond to eigenvalues of the master $T$-operator
and study dynamics of their zeros as functions of the
spectral parameter. This implies a remarkable
connection between
the quantum spin chain and the classical Ruijsenaars--Schneider
system of particles.}


\Classification{37K10; 81Q80; 05E05}

\renewcommand{\thefootnote}{\arabic{footnote}}
\setcounter{footnote}{0}

\section{Introduction}

The master $T$-operator was
introduced in \cite{AKLTZ11} (in a preliminary form, it
was previously discussed in \cite{KLT10}).
It is a generating function
for commuting conserved quantities of quantum spin chains and
associated integrable vertex models
which unif\/ies the transfer matrices
on all levels of the nested Bethe ansatz and Baxter's
$Q$-operators in one commuting family.

It was also proven in \cite{AKLTZ11} that the master $T$-operator,
as a function of inf\/initely many auxiliary parameters (``times''),
one of which being the quantum spectral parameter,
satisf\/ies the same hierarchy of bilinear Hirota equations
as the classical tau-function does. This means that any eigenvalue
of the master $T$-operator is a tau-function of a classical
integrable hierarchy.
For f\/inite spin chains with
${\rm GL}(N)$-invariant $R$-matrices this tau-function is a polynomial
in the quantum spectral parameter. The close connection of the spin chain
spectral problem with
integrable many-body systems of classical mechanics comes from
the dynamics of zeros of the polynomial tau-functions.
This is a further development of earlier studies
\cite{KSZ08,KLWZ97,Z97,Z97a,KSZ08a}
clarifying the role of
the Hirota bilinear dif\/ference equation \cite{Hirota81,Miwa82}
in quantum integrable models.

In this paper we review the results of \cite{AKLTZ11} and make
the connection with classical many-body systems more
precise. The presentation
here is deliberately made as close as
possible to that of \cite{ALTZ13}, where a similar correspondence
between the quantum Gaudin model and the classical Calogero--Moser
many-body system was established using the connection of the former
model with the Kadomtsev--Petviashvili (KP) hierarchy.
Similarly to that paper, here
we discuss the correspondence between
integrable systems of dif\/ferent kinds:
\begin{itemize}\itemsep=0pt
\item[(i)] Quantum integrable magnets (spin chains) of $XXX$-type,
\item[(ii)] The classical modif\/ied Kadomtsev--Petviashvili (mKP) hierarchy,
\item[(iii)] The classical Ruijsenaars--Schneider (RS) system of particles.
\end{itemize}
The link (i)-(ii) is
the correspondence between quantum spin chains
with the ${\rm GL}(N)$-invariant
rational $R$-matrices and the classical
mKP hierarchy based on the construction of the master $T$-operator
\cite{AKLTZ11,Z12a,Z12}.
The link (ii)-(iii) is a well-known story about
dynamics of poles of rational solutions to soliton equations, see
\cite{AMM,Iliev,Krichever-rata,Krichever-rat,KZ95,Shiota,Wilson}.
The composition of (i)-(ii) and
(ii)-(iii) implies the connection between the quantum $XXX$-model and the
classical rational RS model \cite{RS} which was f\/irst mentioned in
\cite{AKLTZ11}.
The link (i)-(iii) also extends the
correspondence between the quantum Gaudin model and the classical
Calogero--Moser system earlier established in \cite{MTV,MTVa}
using dif\/ferent arguments.

Following \cite{AKLTZ11}, we show how to construct
the master $T$-operator
for the ${\rm GL}(N)$-based inhomogeneous
spin chain with twisted boundary conditions
using the co-derivative operation \cite{Kazakov:2007na}.
We call such models ``spin chains'' in a rather broad sense,
not implying the existence of any local Hamiltonian
of Heisenberg type.
(Integrable local interactions in general do
not exist for inhomogeneous spin chains.)
However, even in the general case of arbitrary
inhomogeneity parameters $u_i$
the model still makes sense as a generalized spin chain with
non-local interactions. The ``spin variables'' are
vectors from the spaces ${\mathbb C} ^N$ at each site.
In fact
one may prefer to keep in mind integrable lattice models of statistical
mechanics rather than spin chains as such.
In either case the f\/inal goal of the theory is the diagonalization of the
transfer matrices which is usually achieved by the nested
Bethe ansatz method in one form or another.

The master $T$-operator depends on an inf\/inite number
of auxiliary ``time variables'' ${\bf t}=\{t_0, t_1, t_2, \ldots \}$
(where $t_0$ can be identif\/ied with the spectral parameter $u$)
and satisf\/ies the bilinear identity
for the classical mKP hierarchy. Hence
any of its eigenvalues is a mKP tau-function.
Here is a short dictionary of the $XXX$-mKP correspondence:

\begin{center}
\begin{tabular}{ccc}
\bf $\boldsymbol{XXX}$ chain & & \bf mKP hierarchy \\[7pt]
{\it master $T$-operator} & $\longleftrightarrow$ &
{\it $\tau$-function}\\[4pt]
{\it spectral parameter} & $\longleftrightarrow$ &
{\it the $t_0$-variable}\\[4pt]
\it higher transfer matrices& $\longleftrightarrow$ &
\it Pl\"ucker coordinates
\end{tabular}
\end{center}

\noindent
Moreover, from the explicit
form of the $R$-matrix and the Yang--Baxter equation it follows that
this tau-function is a polynomial in $u=t_0$. Therefore, according to
\cite{KZ95,Iliev}, the dynamics of its
roots in $t_i$ with $i\geq 1$ is
given by equations of motion of the rational RS system
of particles. It should be mentioned that the method for deriving
the dynamics of roots is similar to that used in deriving the
Bethe equations in Sklyanin's
separation of variables method~\cite{Sklyanin,Sklyanina}.
This is not particularly surprising because, according to~\cite{NRK} (see also~\cite{KLWZ97,Z97}), the nested Bethe ansatz
equations themselves can be understood as an integrable
many-body system of RS type in discrete time.

The $XXX$-RS correspondence implies that
the ``inhomogeneities at sites'' $u_i$ in the $XXX$-chain
should be identif\/ied with
initial coordinates of the RS particles while eigenvalues
of the spin chain Hamiltonians are their initial velocities.
Eigenvalues of the Lax matrix for the rational RS model coincide with
eigenvalues of the twist matrix (with certain multiplicities).
Therefore, with f\/ixed integrals of motion
in the RS model determined by invariants of the
twist matrix, there are a f\/inite number of solutions for their values
which correspond to dif\/ferent eigenstates of the spin chain.
In other words, the eigenstates of the spin chain Hamiltonians
are in one-to-one correspondence with (a f\/inite number of)
intersection points
of two Lagrangian submanifolds in the phase space of the classical RS model.
One of them is the hyperplane of f\/ixed $u_i$'s and another
is the submanifold of constant levels of the RS Hamiltonians
in involution.
In short, the dictionary of the $XXX$-RS correspondence is
as follows:

\begin{center}
\begin{tabular}{ccc}
\bf $\boldsymbol{XXX}$ chain & & \bf Ruijsenaars--Schneider \\[7pt]
{\it inhomogeneities at the sites} & $\longleftrightarrow$ &
{\it initial
coordinates }\\[4pt]
\it eigenvalues of Hamiltonians & $\longleftrightarrow$ &
\it initial momenta \\[4pt]
\it twist parameters  & $\longleftrightarrow$ &
\it integrals of motion
\end{tabular}
\end{center}

\noindent
This ``quantum-classical correspondence''
was also discussed \cite{GK13,GZZ, NRS} in the context of
supersymmetric gauge theories and branes.

\section{The quantum spin chain}

Consider generalized quantum integrable spin chains with
${\rm GL}(N)$-invariant $R$-matrix
\[
R(u)=I \otimes I +\frac{1}{u}
\sum_{a,b=1}^{N}e_{ab}\otimes e_{ba}.
\]
Here $u$ is the spectral parameter and
$I$ is the unity matrix.
By $e_{ab}$ we denote the
basis in the space of $N\! \times \! N$ matrices
such that $e_{ab}$ has only
one non-zero element (equal to 1) at the place $ab$:
$(e_{ab})_{cd}=\delta_{ac}\delta_{bd}$. Note that
$P=\sum_{ab}e_{ab} \otimes e_{ba}$
is the permutation matrix
in the space ${\mathbb C}^N \otimes {\mathbb C}^N$ with the def\/ining property
$P(v\otimes w)=w\otimes v$ for any vectors $v,w\in {\mathbb C}^N$, so
the $R$-matrix can be written as
$R(u)=I\otimes I+\frac{1}{u}P$. The ${\rm GL}(N)$-invariance
of this $R$-matrix means that
$g\otimes g \, R(u)=R(u)g\otimes g$ for any $g\in {\rm GL}(N)$.

A more general ${\rm GL}(N)$-invariant $R$-matrix is
\begin{gather}\label{R2}
R_{\lambda}(u)=I \otimes I +\frac{1}{u}\sum_{a,b=1}^{N}
e_{ab}\otimes \pi_{\lambda}({\sf e}_{ba}),
\end{gather}
which acts in the tensor product of the vector representation
space ${\mathbb C}^N$ and an arbitrary f\/inite-dimensional irreducible
representation $\pi_{\lambda}$ of the algebra $U(gl(N))$
with highest weight $\lambda$.
We identify $\lambda$ with the Young diagram $\lambda =
(\lambda_1, \lambda_2 , \ldots , \lambda_{\ell})$ with
$\ell=\ell(\lambda )$ non-zero rows, where
$\lambda_i \in {\mathbb Z}_{+}$,
$\lambda_1 \geq \lambda_2 \geq \cdots \geq \lambda_{\ell}>0$.
By ${\sf e}_{ab}$ we denote the generators of the algebra
$U(gl(N))$ with the commutation relations
${\sf e}_{ab}{\sf e}_{a'b'}-{\sf e}_{a'b'}{\sf e}_{ab}=
\delta_{a'b}{\sf e}_{ab'}-\delta_{ab'}{\sf e}_{a'b}$.
In this notation we have $e_{ab}=\pi_{(1)}({\sf e}_{ab})$,
where $\pi_{(1)}$ is the $N$-dimensional vector representation
corresponding to the 1-box diagram $\lambda =(1)$.

For working in multiple
tensor product spaces like $({\mathbb C}^N)^{\otimes n}=
\underbrace{{\mathbb C}^N \otimes \cdots \otimes {\mathbb C}^N}_{n}$
the following notation is convenient. For any $g\in \mbox{End}\, ({\mathbb C}^N)$
we write $g^{(i)}=I^{\otimes (i-1)} \otimes
g\otimes  I^{\otimes (n-i)}\in \mbox{End}\bigl (({\mathbb C}^N)^{\otimes n}\bigr )$.
In particular, the generators of ${\rm GL}(N)$ can be realized as
$e^{(i)}_{ab}:= I^{\otimes (i-1)} \otimes
e_{ab} \otimes  I^{\otimes (n-i)}$.
They commute for
any $i \ne j$ because they act in dif\/ferent spaces.
In this notation, $ P_{ij}=
P_{ji}=\sum_{a,b} e_{ab}^{(i)}e_{ba}^{(j)} $
($i\neq j$)
is the operator acting by permutation of the $i$-th and $j$-th tensor
factors in the space $({\mathbb C}^N)^{\otimes n}$. We have:
$P_{ij}g^{(j)}=g^{(i)}P_{ij}$,
and $P_{ij}g^{(k)}=g^{(k)}P_{ij}$ for
$k\neq i,j$.

Fix a matrix $g\in {\rm GL}(N)$ called
the twist matrix which is assumed to be diagonalizable.
A~family of commuting operators acting in the space
${\cal V}=({\mathbb C}^N)^{\otimes n}$
(quantum transfer matrices or $T$-operators)
can be constructed as
\begin{gather}\label{R3}
{\sf T}_{\lambda}(u)=\mbox{tr}_{\pi_{\lambda}}\Bigl (
R_{\lambda}^{10}(u-u_1)R_{\lambda}^{20}(u-u_2)
 \cdots  R_{\lambda}^{n0}(u-u_n)
( I^{\otimes n} \otimes \pi_{\lambda}(g) )\Bigr ),
\end{gather}
where $u_i$ are arbitrary complex parameters which are assumed to be
all distinct.
The trace is taken in the auxiliary
space $V_{\lambda}$ where the representation $\pi_{\lambda}$ is realized.
By $R_{\lambda}^{j0}(u)$ we denote
the $R$-matrix \eqref{R2} acting  in the tensor product of
the $j$-th local space~${\mathbb C}^N$ of the chain and the space~$V_{\lambda}$
labeled by~$0$. More precisely,
let us denote~\eqref{R2} symbolically as
$R_{\lambda}(u)=\sum_{i}a_{i}\otimes b_{i} $. Then~$R_{\lambda}^{j0}(u)$ is realized as
$R_{\lambda}^{j0}(u)=\sum_{i} I^{\otimes (j-1)} \otimes a_{i}
 \otimes I^{\otimes (n-j)} \otimes b_{i}$,
where $j=1,2,\dots , n$.
Here the operator $b_{i}$ acts in the auxiliary space
$V_{\lambda}$.
It follows from the Yang--Baxter equation that the $T$-operators
with the same $g$, $u_i$ commute for all $u$, $\lambda$ and can be
simultaneously diagonalized.
The normalization used above is such that
${\sf T}_{\varnothing}(u)=I^{\otimes n}$.
Another useful normalization
is
\[
T_{\lambda}(u)=\prod_{j=1}^{n}(u-u_j) \cdot
{\sf T}_{\lambda}(u).
\]
In this normalization all $T_{\lambda}(u)$ and all their eigenvalues
are polynomials in $u$ of degree $n$.

For $n=0$ the transfer matrix
(\ref{R3}) is the character of $g$ in the representation
$\pi_{\lambda}$:
$
{\sf T}_{\lambda}(u)=\mbox{tr}_{\pi_{\lambda}}g :=\chi_{\lambda}(g)
$,
It is given by the
Schur polynomial $s_{\lambda}({\bf y})$
of the variables ${\bf y}=\{y_1, y_2, \ldots \}$, where
$y_k=\frac{1}{k}\, \mbox{tr}\, g^k$:
\[
\chi_{\lambda}(g)=s_{\lambda}({\bf y})=
\det_{i,j=1, \ldots , \ell (\lambda )}h_{\lambda_i -i+j}({\bf y}),
\]
(the Jacobi--Trudi formula).
Here the complete symmetric polynomials
$h_{k}({\bf y})=s_{(k)}({\bf y})$ are def\/ined by
\[
\exp \bigl (\xi({\bf y},z) \bigr )=
\sum_{k=0}^{\infty}h_{k}({\bf y})z^{k}, \qquad \xi({\bf y},z)
:= \sum_{k\geq 1}y_kz^k.
\]
It is convenient to set $h_{k}=0$ for $k<0$.
Let $w_1, \ldots , w_N$ be the eigenvalues of $g\in {\rm GL}(N)$ realized
as an element of $\mbox{End} \, ({\mathbb C}^N)$. Then
$y_k = \frac{1}{k}\, (w_1^k +\cdots +w_N^k)$ and
\[
\chi_{\lambda}(g)= \frac{\det_{1\leq i,j\leq N}
\bigl (w_j^{\lambda _i +N-i}\bigr )}{\det_{1\leq i,j\leq N}
\bigl (w_j^{N-i}\bigr )}
\]
(see~\cite{Macdonald}). This formula implies that
$\chi_{\varnothing}(g)=s_{\varnothing}({\bf y})=1$.

A more explicit
construction of the quantum transfer matrices
$T_{\lambda}(u)$ was suggested in
\cite{Kazakov:2007na}. It uses the special derivative operator
on the group ${\rm GL}(N)$ called there the co-derivative operator.
In fact it is a sort of ``matrix logarithmic derivative''.
The precise def\/inition is as follows.
Let $g$ be an element of the group ${\rm GL}(N)$ and $f$ be any
function of $g$ with values in $\mbox{End}\, (L)$, where
$L$ is the space of any ${\rm GL}(N)$-representation.
The (left) co-derivative is def\/ined as
\[
{\sf D}f(g)=\frac{\p}{\p \varepsilon} \sum_{ab}e_{ab}\otimes
f\big(e^{\varepsilon {\sf e}_{ba}}g\big)\Bigr |_{\varepsilon =0}.
\]
The right hand side belongs to $\mbox{End} \, ({\mathbb C}^N \otimes L)$.
In particular, the result of the action of~${\sf D}$
on a~scalar function is a~linear
operator in ${\mathbb C}^N$, acting by ${\sf D}$ twice we get an operator
in ${\mathbb C}^N \otimes {\mathbb C}^N$ and so on.
For example:
\begin{gather*}
{\sf D}\det g = \det g \cdot I,\qquad
{\sf D}\, \mbox{tr}\, g^m= mg^m,\qquad
{\sf D}g^m = \sum_{k=0}^{m-1}P \big(g^k\otimes g^{m-k}\big) \quad
\mbox{for $k\geq 1$}.
\end{gather*}
When the number of tensor factors is more than two
another notation is more convenient.
Let $V_i\cong {\mathbb C}^N$ be copies of ${\mathbb C}^N$
and ${\cal V}=V_1\otimes \cdots \otimes V_n$ as before.
Then,
applying the matrix derivatives to a scalar function $f$
several times, we can embed the result into $\mbox{End} ({\cal V})$
according to the formulas
\begin{gather*}
{\sf D}_{i_1}f(g)=
\sum_{a,b}e_{ab}^{(i_1)}f\big(e^{\varepsilon
{\sf e}_{ba}}g\big)\Bigr |_{\varepsilon =0},
\\
{\sf D}_{i_2} {\sf D}_{i_1} f(g)=
\frac{\p}{\p \varepsilon_2}
\frac{\p}{\p \varepsilon_1}
\sum_{a_2b_2}\sum_{a_1b_1}
e^{(i_2)}_{a_2b_2}e^{(i_1)}_{a_1b_1}
f\Big (e^{\varepsilon _1{\sf e}^{(i_1)}_{b_1a_1}}
e^{\varepsilon _2{\sf e}^{(i_2)}_{b_2a_2}}
g\Big)
\Bigr |_{\varepsilon_1=\varepsilon_2=0},
\end{gather*}
and so on. The lower indices of ${\sf D}$ show in which tensor
factors the resulting operator acts non-trivially.
In this notation, the examples given above read:
${\sf D}_i \mbox{tr}\, g=g^{(i)}$,
${\sf D}_i \mbox{tr}\, g^{(j)}=P_{ij}g^{(j)}$ ($i\neq j$).
For many other formulas of this type see
\cite[Appendix~D]{AKLTZ11}.

According to \cite{Kazakov:2007na}
the transfer matrix (\ref{R3}) can be represented as a
chain of operators of the form $1+\frac{{\sf D}_i}{u-u_i}$
acting on the character:
\[
{\sf T}_{\lambda}(u)=\left ( 1+\frac{{\sf D}_n}{u-u_n}\right)
\cdots \left( 1+\frac{{\sf D}_1}{u-u_1}\right)
\chi_{\lambda}(g) .
\]
For simplicity we assume that the twist matrix $g$ is diagonal:
$g=\mbox{diag}\, (w_1, w_2, \ldots , w_N)$.
By analogy with the Gaudin model, one may introduce
(non-local) spin chain ``Hamiltonians'' as residues of
the ${\sf T}_{(1)}(u)$ at $u=u_i$:
\begin{gather}\label{D2a}
{\sf T}_{(1)}(u)=\mbox{tr}\, g + \sum_{i=1}^{n}\frac{H_i}{u-u_i}.
\end{gather}
Explicitly, they have the form:
\[
H_i=\overrightarrow{\prod_{j=i+1}^{n}}\left (
1+\frac{P_{ij}}{u_i-u_j}\right )g^{(i)}
\overrightarrow{\prod_{\,\, j=1\,\,}^{i-1}}\left (
1+\frac{P_{ij}}{u_i-u_j}\right ).
\]
(Here and below
we write
$ {\overrightarrow{\prod\limits_{j=1}^{m}}A_j =
A_1A_2\cdots A_m}$ and
$ {\overleftarrow{\prod\limits_{j=1}^{m}}A_j =
A_m \cdots A_2 A_1}$ for ordered products.) For example, for $n=3$ we have:
\begin{gather*}
H_1=\left ( 1+\frac{P_{12}}{u_1-u_2}\right )
\left ( 1+\frac{P_{13}}{u_1-u_3}\right )g^{(1)},
\\
H_2=\left ( 1+\frac{P_{23}}{u_2-u_3}\right )g^{(2)}
\left ( 1+\frac{P_{21}}{u_2-u_1}\right ),
\\
H_3=g^{(3)}\left ( 1+\frac{P_{31}}{u_3-u_1}\right )
\left ( 1+\frac{P_{32}}{u_3-u_2}\right ) .
\end{gather*}
Similar non-local operators were discussed in \cite{HKW92}.

It is easy to check that the operators
\begin{gather}\label{Ma}
M_a=\sum_{l=1}^{n}e_{aa}^{(l)}
\end{gather}
commute with the
Hamiltonians $H_i$: $[H_i, M_a]=0$ (for diagonal $g$).
Therefore, common eigenstates of the Hamiltonians can be classif\/ied
according to eigenvalues of the operators $M_a$. Let
\[
{\cal V}=\bigotimes_{i=1}^{n}V_i   =
\bigoplus_{m_1, \ldots , m_N}   {\cal V}(\{m_a \})
\]
be the decomposition of the Hilbert space of the spin chain
${\cal V}$ into the direct sum of eigenspaces of the operators
$M_a$ with eigenvalues $m_a \in {\mathbb Z}_{\geq 0}$, $a=1, \ldots , N$.
Then eigenstates of the $H_i$'s belong to the spaces ${\cal V}(\{m_a \})$.
Since $\sum_a e_{aa}=I$ is the unit matrix,
$\sum_a M_a =n I ^{\otimes n}$, and hence
\[
\sum_{a=1}^{N}m_a =n.
\]
Note also that
\[
\sum_{i=1}^{n}H_i= ({\sf D}_1+
\cdots + {\sf D}_n)\, \mbox{tr} \, g =
\sum_{i=1}^ng^{(i)}=
\sum_{i=1}^n \sum_{a=1}^N e_{aa}^{(i)}w_a=\sum_{a=1}^{N}w_a M_a .
\]

\section[The master $T$-operator and the mKP hierarchy]{The master $\boldsymbol{T}$-operator and the mKP hierarchy}

\subsection[The master $T$-operator]{The master $\boldsymbol{T}$-operator}

The master $T$-operator for the spin chain can be def\/ined as
\begin{gather}\label{master1}
T(u, {\bf t})=(u-u_n +{\sf D}_n)  \cdots
(u-u_1 +{\sf D}_1)\exp \left (\sum_{k\geq 1}t_k \, \mbox{tr}\, g^k\right),
\end{gather}
where ${\bf t}=\{t_1, t_2, \ldots \}$ is an inf\/inite set of ``time
parameters''. These operators commute for all values of the parameters:
$[T(u,{\bf t}),   T(u',{\bf t'})]=0$.

The Cauchy--Littlewood identity
\[
\sum_{\lambda}\chi_{\lambda}(g)s_{\lambda}({\bf t})=
\exp \left(\sum_{k\geq 1}t_k \, \mbox{tr}\, g^k\right)
\]
implies that the
expansion of $T(u, {\bf t})$ in the Schur functions is
\begin{gather}\label{master2}
T(u, {\bf t})=\sum_{\lambda}T_{\lambda}(u)s_{\lambda}({\bf t}).
\end{gather}
The sum is taken over all Young diagrams including
the empty one. Therefore,
the $T$-opera\-tors~$T_{\lambda}(u)$ can be restored from the master
$T$-operator according to the formula
\begin{gather}\label{master3}
T_{\lambda}(u)=s_{\lambda}(\tilde \p )T(u, {\bf t})\Bigr |_{{\bf t}=0},
\end{gather}
where $\tilde \p =\{ \p_{t_1}, \frac{1}{2}\, \p_{t_2},
\frac{1}{3}\, \p_{t_3}, \ldots \}$.
In particular,
\[
T_{(1)}(u)=\p_{t_1}T(u, {\bf t})\Bigr |_{{\bf t}=0}, \qquad
T_{(1^2)}(u)=\frac{1}{2}
\big(\p_{t_1}^2 -\p_{t_2}\big)T(u, {\bf t})\Bigr |_{{\bf t}=0}.
\]

Given $z\in {\mathbb C}$, we will use the standard
notation ${\bf t}\pm [z^{-1}]$ for the following special
shift of the time variables:
\[
{\bf t}\pm [z^{-1}]:= \left\{t_1 \pm z^{-1},   t_2 \pm \frac{1}{2}
z^{-2},   t_3 \pm \frac{1}{3}
z^{-3},   \ldots \right \}.
\]
As we shall see below,
$T(u, {\bf t}\pm [z^{-1}])$ regarded as functions of
$z$ with f\/ixed ${\bf t}$ plays an important role.
Here we only note that equation~(\ref{master3}) implies that
$T(u, 0\pm [z^{-1}])$ is the generating series for
$T$-operators corresponding to the one-row and one-column diagrams
respectively:
\begin{gather}\label{master4}
T\big(u,  \big[z^{-1}\big]\big)=\sum_{s\geq 0}z^{-s}T_{(s)}(u), \qquad
T\big(u,  -\big[z^{-1}\big]\big)=\sum_{a=0}^{N}(-z)^{-a}T_{(1^a)}(u).
\end{gather}

\subsection{The bilinear identity and Hirota
equations}
\label{sec-limitproof}

The following statement was proved in
\cite{AKLTZ11}:

\begin{Theorem}
The master $T$-operator \eqref{master1}
satisfies the bilinear identity for the mKP hierarchy
{\rm \cite{DJKM83,JM83}}:
\begin{gather}\label{hir1}
\oint_{{\cal C}_{[0,\infty ]}}
z^{u-u'}e^{\xi ({\bf t}-{\bf t'}, z)}
T\left (u, {\bf t}-[z^{-1}]\right )
T\left (u, {\bf t'}+[z^{-1}]\right ){\rm d}z =0
\end{gather}
for all ${\bf t}$, ${\bf t}'$ and $u$, $u'$, where
the integration contour ${\cal C}_{[0,\infty ]}$
encircles the cut
$[0,\infty ]$ between~$0$ and~$\infty $
$($including the points~$0$ and~$\infty)$ and does not enclose
any singularities coming from the $T$-factors.
\end{Theorem}

This means that each eigenvalue of the master $T$-operator
is a tau-function of the mKP hierarchy.
Equation~(\ref{master2}) is the expansion of the tau-function
in Schur polynomials~\cite{EH, Orlov-Shiota, Sato}.
The functional relations for quantum transfer matrices
\cite{BR90,Chered,KNS,KNSa} can be then interpreted as
Pl\"ucker relations for coef\/f\/icients of the expansion.

Setting $u'=u$ and
$t_k'=t_k-\frac{1}{k}\big(z_{1}^{-k}+
z_{2}^{-k}+z_{3}^{-k}\big)$ in~(\ref{hir1})
and taking the residues
we arrive at the 3-term Hirota equation
\begin{gather}
(z_2-z_3)T\left (u, {\bf t}-\big[z_{1}^{-1}\big]\right )T
\left (u, {\bf t}-\big[z_{2}^{-1}\big]-\big[z_{3}^{-1}\big]\right )
\nonumber\\
\qquad{}+
(z_3-z_1)T\left (u, {\bf t}-\big[z_{2}^{-1}\big]\right )T
\left (u, {\bf t}-\big[z_{3}^{-1}\big]-\big[z_{1}^{-1}\big]\right )
\nonumber \\
\qquad{} +  (z_1-z_2)T\left (u, {\bf t}-\big[z_{3}^{-1}\big]\right )T
\left (u, {\bf t}-\big[z_{1}^{-1}\big]-\big[z_{2}^{-1}\big]\right )   =  0.\label{bi2}
\end{gather}
Setting $u'=u-1$,
$t_k'=t_k-\frac{1}{k}\big(z_{1}^{-k}+
z_{2}^{-k}\big)$,
we obtain another 3-term Hirota equation
\begin{gather*}
z_2T\left (u+1,{\bf t}-\big[z_{2}^{-1}\big]\right )
T\left (u,{\bf t}-\big[z_{1}^{-1}\big]\right )-
z_1 T\left (u+1, {\bf t}-\big[z_{1}^{-1}\big]\right )
T\left (u, {\bf t}-\big[z_{2}^{-1}\big]\right )
\\
\qquad{} + (z_1-z_2)T(u+1, {\bf t})T
\left (u, {\bf t}-\big[z_{1}^{-1}\big]-\big[z_{2}^{-1}\big]\right )  =0.
\end{gather*}
Due to (\ref{BA2b}) (see below), it can be formally regarded
as a particular case of~(\ref{bi2}) in the limit $z_3\to 0$.

\subsection[The Baker-Akhiezer functions]{The Baker--Akhiezer functions}

According to the general scheme, the Baker--Akhiezer (BA)
function and its adjoint correspon\-ding to the tau-function~(\ref{master1})
are given by the formulas \cite{DJKM83,JM83}
\begin{gather}\label{BA1}
\psi_u ({\bf t};z)=z^u e^{\xi ({\bf t}, z)}
T^{-1}(u, {\bf t})  T\bigl (u, {\bf t}-\big[z^{-1}\big]\bigr ),
\\
\label{BA2}
\psi^{*}_u ({\bf t};z)=z^{-u}e^{-\xi ({\bf t}, z)}
T^{-1}(u, {\bf t})  T\bigl (u, {\bf t}+\big[z^{-1}\big]\bigr ).
\end{gather}
For brevity, we will refer to both $\psi$ and $\psi^*$ as
BA functions. In terms of the BA functions,
the bilinear identity (\ref{hir1}) can be written as
\[
\oint_{{\cal C}_{[0,\infty ]}} \psi _u ({\bf t};z)
\psi^{*}_{u'} ({\bf t'};z) {\rm d}z =0.
\]

Using the def\/inition (\ref{master1}), we have:
\begin{gather*}
T\bigl (u, {\bf t}-\big[z^{-1}\big]\bigr ) =
z^{-N}\bigl ( u-u_n +{\sf D}_n\bigr )\cdots \bigl ( u-u_1 +{\sf D}_1\bigr )
\left [ \det (zI-g)e^{{\rm tr}\, \xi ({\bf t},g)}\right],
\\
T\bigl (u, {\bf t}+\big[z^{-1}\big]\bigr )=
z^{N}\bigl ( u-u_n +{\sf D}_n\bigr )\cdots \bigl ( u-u_1 +{\sf D}_1\bigr )
\left [ \frac{ e^{{\rm tr}\, \xi ({\bf t},g)}}{\det (zI-g)}\right ].
\end{gather*}
Note that because $(\det g)^{-1}{\sf D} \det g = {\sf D}+1$, we have
\begin{gather}\label{BA2b}
\lim_{z\to 0}\left ( z^{\pm N}T\big(u,{\bf t}\mp \big[z^{-1}\big]\big)\right )=
(\det g)^{\pm 1}T(u\pm 1, {\bf t}).
\end{gather}
For the BA functions we can thus write:
\begin{gather}\label{BA3}
\psi_{u} ({\bf t};z)=z^{u-N}e^{\xi ({\bf t}, z)}
T^{-1}(u, {\bf t})
\overleftarrow{\prod_{i=1}^{n}}
(u  -  u_i   +  {\sf D}_i)
\left [ \det \left (z I  -  g\right )e^{{\rm tr}\, \xi ({\bf t},g)}
\right ],
\\
\label{BA4}
\psi ^* _{u}({\bf t};z)=z^{N-u} e^{-\xi ({\bf t}, z)}
T^{-1}(u, {\bf t}) \,
\overleftarrow{\prod_{i=1}^{n}}
(u  -  u_i   +  {\sf D}_i)
\left [ \frac{e^{{\rm tr}\,
\xi ({\bf t},g)}}{\det \left (zI  -  g\right )}
\right ].
\end{gather}
From these formulas we see that
$z^{-u}e^{-\xi ({\bf t}, z)}\psi_u ({\bf t};z)$ is a polynomial
in $z^{-1}$ of degree $N$ while
$z^u e^{\xi ({\bf t}, z)}\psi ^*_u ({\bf t};z)$ is a rational
function of $z$ with poles at the points $z=w_i$ (eigenvalues
of the matrix~$g$) of at least f\/irst order because of
$\det (zI-g)$ in the denominator. Moreover, since each co-derivative
raises the order of the poles, these poles may be actually of a higher
order, up to $n+1$. Also, as is seen from the
second formula, this function has a zero of order $N$
at $z=0$. (We assume that $w_a\neq 0$.)

Regarded as functions of $u$, both $z^{-u}\psi_u$ and
$z^u \psi^*_u$ are rational functions of $u$ with
$n$ zeros and $n$ poles which are simple in general position.
From (\ref{master1}) and (\ref{BA3}), (\ref{BA4}) it follows that
\begin{gather} \label{BA5}
\lim_{u\to \infty}\left (z^{-u}e^{-\xi ({\bf t}, z)}\psi_u ({\bf t};z)
\right)=z^{-N}\det (zI-g),
\\
\lim_{u\to \infty}\left (z^u e^{\xi ({\bf t}, z)}\psi ^*_u ({\bf t};z)
\right)=z^{N}(\det (zI-g))^{-1}.\nonumber
\end{gather}

The BA functions satisfy the following dif\/ferential-dif\/ference
equations:
\begin{gather}\label{BA7}
\p_{t_1}\psi_u ({\bf t};z)=\psi_{u+1} ({\bf t};z) +
V(u, {\bf t}) \psi_u ({\bf t};z),
\\
\label{BA8}
-\p_{t_1}\psi ^*_u ({\bf t};z)=\psi ^*_{u-1} ({\bf t};z) +
V(u-1, {\bf t}) \psi ^*_u ({\bf t};z),
\end{gather}
where
\begin{gather}\label{BA9}
V(u, {\bf t})=\p_{t_1} \log \frac{T(u+1, {\bf t})}{T(u, {\bf t})}.
\end{gather}

We also note the formulas for the {\it stationary} BA
functions $\psi_u (z):=\psi_u (0;z)$, $\psi^*_u (z):=\psi^*_u (0;z)$
which directly follow from (\ref{BA3}), (\ref{BA4}):
\begin{gather}\label{BA3st}
\psi_u (z)=z^{u-N}
\overleftarrow{\prod_{i=1}^{n}}
\left (1+\frac{{\sf D}_i}{u-u_i}\right )
\det (zI-g),
\\
\psi ^*_u(z)=z^{N-u}
\overleftarrow{\prod_{i=1}^{n}}
\left(1+\frac{{\sf D}_i}{u-u_i}\right)
\frac{1}{\det (zI-g)}.\nonumber
\end{gather}

Below we will also need the relation
\begin{gather}\label{BA10}
\p_{t_m}\log \frac{T (u+1, {\bf t})}{T (u, {\bf t})}=
\mbox{res}_{\infty}\bigl (\psi_u ({\bf t};z\bigr )
\psi^*_{u+1} ({\bf t};z)z^m \, {\rm d}z \bigr ).
\end{gather}
(Here $\mbox{res}_{\infty}(\ldots )\equiv \frac{1}{2\pi i}
\oint_{\infty}(\ldots )$ and $ \frac{1}{2\pi i}
\oint_{\infty}z^{-1}{\rm d}z =1$.)
This relation can be derived from the bilinear identity (\ref{hir1})
in the following way.
Applying $\p_{t'_m}$ and putting $u'=u+1$, $t'_k=t_k$ afterwards, we get:
\begin{gather*}
\begin{split}
&\oint_{{\cal C}_{[0,\infty ]}}
\Bigl ({-}z^{m-1}
T\left (u, {\bf t}-\big[z^{-1}\big]\right )
T\left (u+1, {\bf t}+\big[z^{-1}\big]\right )
\\
& \qquad{} +  z^{-1}
T\left (u, {\bf t}-\big[z^{-1}\big]\right )\p_{t_m}
T\left (u+1, {\bf t}+\big[z^{-1}\big]\right )
\Bigr ){\rm d}z =0.
\end{split}
\end{gather*}
The f\/irst term is regular at $z=0$ and thus contributes
to the integral by the residue at $\infty$ while the second
term has residues at the points~$0$,~$\infty$ and the both contribute
to the integral. Using~(\ref{BA2b}), we can f\/ind the residues:
\begin{gather*}
\mbox{res}_{\infty}\Bigl ( z^{m-1}T\left (u, {\bf t}-\big[z^{-1}\big]\right )
T\left (u+1, {\bf t}+\big[z^{-1}\big]\right )\Bigr )
\\
\qquad{} =  T(u, {\bf t}) \p_{t_m}T(u+1, {\bf t})-
T(u+1, {\bf t}) \p_{t_m}T(u, {\bf t})
\end{gather*}
Dividing both sides by $T(u, {\bf t})T(u+1, {\bf t})$, we obtain
(\ref{BA10}).

\section[Zeros of the master $T$-operator as Ruijsenaars-Schneider
particles]{Zeros of the master $\boldsymbol{T}$-operator\\ as Ruijsenaars--Schneider
particles}

The eigenvalues of the master $T$-operator are polynomials in
the spectral parameter $u$:
\[
T(u, {\bf t})=e^{t_1 \,  \text{tr}\, g
+t_2 \, \text{tr} \, g^2 +\cdots}
\prod_{k=1}^{n}(u -u_k(t_1, t_2, \ldots )).
\]
The roots of each eigenvalue have their own
dynamics in the times $t_k$. These dynamics are known
to be given by the rational RS model~\cite{RS} (see
\cite{Iliev,KZ95},   which extend the methods
developed by Krichever \cite{Krichever-rat} and Shiota~\cite{Shiota}
for dymamics of poles of solutions to the KP hierarchy).
The inhomogeneity parameters of the spin chain play the role of
coordinates of the RS particles at $t_i=0$:  $u_j=u_j(0)$.
In particular, we have
$T(u,0)=T_{\varnothing}(u)=\prod\limits_{k=1}^{n}(u-u_k)$.

From (\ref{master4}) we see that
$T(u):=T_{(1)}(u)=\p_{t_1}T(u, {\bf t})\bigr |_{{\bf t}=0}$.
Therefore,
\[
{\sf T}_{(1)}(u)= \frac{T_{(1)}(u)}{T_{\varnothing}(u)}=\p_{t_1}
\log T(u, {\bf t})\bigr |_{{\bf t}=0}=
\mbox{tr} \, g   -   \sum_{k=1}^{n}\frac{\dot u_k(0)}{u-u_k}.
\]
Comparing with (\ref{D2a}),
we conclude that
the initial velocities are equal (up to sign) to the eigenvalues of the
spin chain Hamiltonians:
\begin{gather}\label{CM2}
\dot u_i =-H_i.
\end{gather}
This unexpected connection between the quantum spin chain
and the classical RS model was mentioned in \cite{AKLTZ11}.
A similar relation between quantum Hamiltonians
in Gaudin model and velo\-ci\-ties of particles in the classical
Calogero--Moser model was found in \cite{MTV,MTVa} within a~dif\/ferent
framework, see also \cite{MTV1,MTV2} for further developments.

\subsection{Lax pair for the RS model from dynamics of poles}

Following Krichever's method
\cite{Krichever-rat}, let us derive equations of motion
for the $t_1$-dynamics of the~$u_i$'s. Essentially, the derivation below is
not specif\/ic to the master $T$-operator case but only depends on
the polynomiality of the tau-function. The specif\/ic part is the
particular normalization of the BA functions.

It is convenient to denote
$t_1=t$ and put all other times to zero since they are irrelevant for
this derivation. Correspondingly, we will write $T(u,t)$ instead of
$T(u, {\bf t})$ and $\p_t u_k =\dot u_k$, etc.
From~(\ref{BA9}) we see that
\[
V(u,t)=\p_t \log \frac{T(u+1,t)}{T(u,t)}
=\sum_{k=1}^{n}\left (\frac{\dot u_k}{u-u_k}-
\frac{\dot u_k}{u-u_k+1}\right ).
\]
The method of \cite{Krichever-rat} is to perform the pole expansion
of the linear problem (\ref{BA7}) for the BA function $\psi$. From
(\ref{BA3}) we have the pole expansion of the BA function
\[
\psi = z^u e^{tz}\left (c_0 (z) +
\sum_{i=1}^{n}\frac{c_i(z,t)}{u-u_i (t)}\right ),
\]
where $c_0(z)=\det (I -z^{-1}g)$ (see (\ref{BA5})).
Substituting this into (\ref{BA7}), we obtain
\begin{gather*}
\sum_{i=1}^{n}\left ( \frac{zc_i +\dot c_i}{u-u_i}
+\frac{c_i \dot u_i}{(u-u_i)^2}\right )- \sum_{i=1}^{n}\frac{zc_i - c_0 \dot u_i}{u-u_i+1}
\\
\qquad{} - \sum_{i=1}^{n}
\frac{c_0 \dot u_i}{u-u_i}
-\sum_{i=1}^{n}\frac{c_i}{u-u_i}\sum_{k=1}^{n}
\left ( \frac{\dot u_k}{u-u_k}-
\frac{\dot u_k}{u-u_k+1}\right )=0.
\end{gather*}
The l.h.s.\ is a rational function of $u$ with f\/irst
order poles at $u=u_i$ and $u=u_i-1$ (possible
poles of the second order cancel automatically) vanishing at inf\/inity.
Therefore, to solve the linear problem it is enough to cancel
all the poles. Representing the l.h.s.\
as a sum of simple pole terms and equating the coef\/f\/icients
in front of each pole
to zero, we get the following system of equations for
$i=1, \ldots , n$:
\begin{gather*}
zc_i -c_0 \dot u_i -\dot u_i \sum_{k=1}^n\frac{c_k}{u_i-u_k-1}=0,
\\
 \dot c_i  +zc_i -c_0 \dot u_i-
c_i \sum_{k\neq i}\frac{\dot u_k}{u_i-u_k}-
\dot u_i \sum_{k\neq i}\frac{c_k}{u_i-u_k} +c_i
\sum_{k=1}^{n}\frac{\dot u_k}{u_i  -  u_k  +  1}=0.
\end{gather*}
These equations can be rewritten in the matrix form:
\begin{gather}
(zI -Y){\sf c}=c_0 (z)\dot U{\sf 1},\nonumber\\
{\sf \dot c}=T{\sf c},\label{CM5}
\end{gather}
where ${\sf c}=(c_1, c_2, \ldots , c_n)^{\sf t}$,
${\sf 1} =(1, 1, \ldots , 1)^{\sf t}$ are $n$-component
vectors and the $n \! \times \!n$ matrices
$U=U(t)$,
$Y=Y(t)$, $T=T(t)$ are
given by
\begin{gather} \label{CM6}
U_{ij}=u_i \delta_{ij}, \qquad
Y_{ij}= \frac{\dot u_i}{u_i -u_j -1},
\\
T_{ij}=
\left (
\sum_{k\neq i}\frac{\dot u_k}{u_i -u_k}   -
 \sum_{k\neq i} \frac{\dot u_k}{u_i   -  u_k   +  1}
\right )\delta_{ij}
+\left (\frac{\dot u_i}{u_i   -  u_j}-
\frac{\dot u_i}{u_i   -  u_j   -  1}\right )
  (1  -  \delta_{ij} ).\nonumber
\end{gather}
Note that $Y=\dot U Q$, $T=\tilde T -Y$, where
\begin{gather}\label{Q}
Q_{ij}=\frac{1}{u_i-u_j-1} ,
\\
\label{tildeT}
\tilde T_{ij}=
\left (
\sum_{k\neq i}\frac{\dot u_k}{u_i -u_k}   -
 \sum_{k} \frac{\dot u_k}{u_i   -  u_k   +  1}
\right )\delta_{ij}+\frac{\dot u_i}{u_i-u_j}
\bigl (1  -  \delta_{ij}\bigr ).
\end{gather}

The compatibility condition of the system (\ref{CM5}) is
$
  ([T,Y]-\dot Y ){\sf c}=c_0 (\ddot U-T\dot U )
{\sf 1}
$ or
\[
- (\ddot UQ +M ){\sf c}=c_0 (\ddot U-T\dot U  ){\sf 1},
\]
where $M:=\dot U\big ( \dot Q +QT -\dot U^{-1}T \dot U Q \big)$.
A straightforward calculation shows that $M=WQ$, where $W$ is the
diagonal matrix $W=\mbox{diag}(W_1, \ldots , W_n)$ with elements
\[
W_i= \sum_{k\neq i}\frac{2\dot u_i
\dot u_k}{(u_i  -  u_k)((u_i  -  u_k)^2-1)}.
\]
Therefore, $[T,Y]-\dot Y=-(\ddot UQ +M)=-(\ddot U +W)Q$. Since
$Q$ is a non-degenerate matrix, the matrix equation
$[T,Y]-\dot Y=0$ is equivalent to $\ddot U +W=0$. At the same time
one can easily check that
\[
 (T\dot U  {\sf 1}  )_i=\sum_{k}T_{ik}\dot u_k = -W_i
\]
and so the compatibility condition for the linear system (\ref{CM5}) is
$\ddot U +W=0$ which yields the equations of motion for the
RS model with $n$ particles
\begin{gather}\label{CM9}
\ddot u_i = -\sum_{k\neq i}
\frac{2\dot u_i \dot u_k}{  (u_i   - u_k  )
 ((u_i  -  u_k)^2-1 )} , \qquad   i=1, \ldots , n.
\end{gather}
Their derivation implies that they can be represented
in the Lax form
\begin{gather}\label{CM8}
\dot Y=[T,   Y].
\end{gather}
and the matrices $Y$, $T$ form the Lax pair for the model.
The matrix $Y$ is the Lax matrix for the RS model.
As is seen from (\ref{CM8}), the time evolution preserves
its spectrum, i.e., the coef\/f\/icients~${\cal J}_k$ of the characteristic polynomial
\[
\det (zI -Y(t))=\sum_{k=0}^{n}{\cal J}_k z^{n-k}
\]
are integrals of motion.

In a similar way, substituting the adjoint BA function
\[
\psi ^*= z^{-u}e^{-tz}\left (c_0^{-1} (z) +
\sum_{i=1}^{n}\frac{c_i^*(z,t)}{u-u_i (t)}\right )
\]
into the adjoint linear problem (\ref{BA8}), we get
\begin{gather}
{\sf c^{*t}}\dot U^{-1}
(zI -Y)=-c_0^{-1} (z){\sf 1^t},\nonumber
\\
\p_t({\sf c^{*t}}\dot U^{-1})=-{\sf c^{*t}}\dot U^{-1}T,\label{CM5a}
 \end{gather}
where ${\sf c^{*t}}=(c_1^*, c_2^*, \ldots , c_n^*)$ and
${\sf 1^t}=(1, 1, \ldots , 1)$. Note that ${\sf 1^t}T=0$.

Using (\ref{CM5}), (\ref{CM5a}), we f\/ind the solutions for the vectors
${\sf c}$, ${\sf c}^*$:
\begin{gather}
{\sf c}(z,t)=c_0(z)(zI -Y(t))^{-1}\dot U{\sf 1} ,\nonumber
\\
{\sf c^{*t}}(z,t)=-c_0^{-1}(z){\sf 1^t}(zI -Y(t))^{-1}\dot U.\label{cc}
\end{gather}
For the functions $\psi$, $\psi^*$ themselves we then have:
\begin{gather}
\psi =c_0(z)  z^u e^{tz}\bigl (
1+{\sf 1}^{\sf t}(uI -U(t))^{-1}(zI -Y(t))^{-1}\dot U{\sf 1}\bigr ) ,
\nonumber\\
 \psi^*=c_0^{-1}(z)z^{-u}e^{-tz}\bigl (
1-{\sf 1}^{\sf t}(zI -Y(t))^{-1}(uI -U(t))^{-1}\dot U{\sf 1}\bigr ) .\label{cc1}
\end{gather}

Let us mention some properties of the matrices $U$, $Y$
to be used in the calculations below.
As is well known (and easy to check), the matrix
$[U,   Y]-Y$ has rank~1. More precisely,
the matrices $U$, $Y$ satisfy the commutation
relation
\begin{gather}\label{comm}
[U,   Y]=Y +U  {\sf 1}\otimes {\sf 1}^{\sf t}
\end{gather}
(here ${\sf 1}\otimes {\sf 1}^{\sf t}$ is the $n\! \times \! n$
matrix of rank $1$ with all entries equal to $1$).

\begin{Lemma}\label{lemma4.1} For any $k\geq 0$ the following equality holds:
\[
{\sf 1}^{\sf t} Y^k \dot U {\sf 1} = -\mbox{{\rm tr}}\, Y^{k+1}.
\]
\end{Lemma}

Indeed, we have: ${\sf 1}^{\sf t} Y^k \dot U{\sf 1}=
\mbox{tr}\, \bigl ({\sf 1}\otimes {\sf 1}^{\sf t}  Y^k\dot U
\bigr )=
\mbox{tr}\, \bigl ((\dot U{\sf 1}\otimes {\sf 1}^{\sf t})  Y^k\bigr )=
\mbox{tr}\, \bigl (([U,   Y]-Y)Y^k\bigr )=-\mbox{tr}\, Y^{k+1}+
\mbox{tr}\, [U,   Y^{k+1}]$ but the last trace
is $0$ as trace of a commutator.

\subsection{Eigenvalues of the Lax matrix}

Here we prove that the eigenvalues of the Lax matrix $Y$
are the same as the eigenvalues of the twist matrix $g$ with appropriate
multiplicities.

\begin{Theorem}
The Lax matrix $Y$ has
eigenvalues $w_a$ with multiplicities $m_a\geq 0$ such that
$m_1 +\cdots +m_N =n$.
\end{Theorem}

Indeed, let us compare the large $|u|$
expansions of (\ref{BA3st}) and
(\ref{cc1}). From (\ref{BA3st}) we have:
\[
\psi_u (z)=\det \big(I-z^{-1}g\big) z^u \left (
1-\frac{1}{u}\sum_i \sum_a \frac{e_{aa}^{(i)}w_a}{z-w_a} +
O\big(u^{-2}\big)\right ).
\]
The expansion of (\ref{cc1}) at $t=0$ gives (using Lemma~\ref{lemma4.1}):
\[
\psi_u (z)=\det \big(I-z^{-1}g\big) z^u \left (
1-\frac{1}{u}\, \mbox{tr} \,\frac{Y_0}{zI-Y_0}   +O\big(u^{-2}\big)\right ),
\]
where we set $Y_0:=Y(0)$.
Therefore, we conclude that
\[
\mbox{tr} \,\frac{Y_0}{zI-Y_0} =
\sum_i \sum_a \frac{e_{aa}^{(i)}w_a}{z-w_a}
\]
and, since $\mbox{tr}\, (zI-Y_0)^{-1}=\p_z \log \det (zI-Y_0)$, we have
\[
\det (zI-Y_0)=\prod_{a=1}^{N}(z-w_a)^{\sum\limits_{i=1}^{n}e_{aa}^{(i)}}
=\prod_{a=1}^{N}(z-w_a)^{M_a},
\]
where $M_a$ is the operator~(\ref{Ma}). Hence we see that
the~$M_a$ is the ``operator multiplicity'' of the
eigenvalue~$w_a$. In the sector~${\cal V}(\{m_a\})$ the
multiplicity becomes equal to~$m_a$.

Less formal arguments are as follows.
The singularities of the vectors
${\sf c}(z,t)$, ${\sf c}^*(z,t)$ as functions
of $z$ are the same as the singularities of the functions
$\psi$, $\psi^*$ in the f\/inite part of the complex plane.
From (\ref{BA3}) we see that ${\sf c}(z,t)$ has a pole of order
$N$ at $z=0$ and no other poles. At the same time the f\/irst equation in~(\ref{cc}) states that there are possible poles at eigenvalues of
the matrix $Y(t)$ (which do not depend on time).
Therefore, they must be canceled by zeros
of $c_0(z)=z^{-N}\det (zI -g)$
which are at $z=w_a$ and are assumed to be simple.
If all eigenvalues of $Y$ are distinct, such a cancellation
is only possible if $n\leq N$. However, the most interesting
setting for the quantum spin chains is quite opposite: $n>N$
or even $n\gg N$
(large chain length at a f\/ixed rank of the symmetry algebra).
We conclude that in
this case the Lax matrix has to have multiple eigenvalues.
At f\/irst glance, a multiple eigenvalue~$w_a$
with multiplicity $m_a\geq 2$
might lead to an unwanted pole of~$\psi$ at $z=w_a$
coming from the higher order pole of
the matrix $(zI - Y)^{-1}$ which now
can not be cancelled by the {\it simple} zero of $\det (zI -g)$.
In fact higher order poles do not appear in the vector
$(zI - Y)^{-1}{\sf 1}$ because $\dot U{\sf 1}$
is a special vector for the matrix~$Y$ which
can be decomposed into~$N$
Jordan blocks of sizes $m_a\! \times \! m_a$.
However, they do appear in the co-vector
${\sf 1^t}(zI - Y)^{-1}$ and the function~$\psi^*$ has
multiple poles at~$z=w_a$ (with multiplicities~$m_a+1$).

\subsection{Equations of motion in Hamiltonian form}

The momenta $v_i$ canonically conjugate to the coordinates
of the RS particles $u_i$ can be introduced by the formula
\[
\dot u_i =-e^{-v_i}\prod_{k\neq i}
\frac{u_i  -  u_k   +  1}{u_i -u_k}
\qquad
\text{or}
\qquad
v_i=-\log (-\dot u_i)+\sum_{k\neq i}\log
\frac{u_i   -  u_k   +  1}{u_i -u_k} .
\]
Then the Hamiltonian form of the RS equations of motion (\ref{CM9})
is
\[
 \begin{pmatrix}\dot u_i \\ \dot v_i \end{pmatrix} =
 \begin{pmatrix} \p_{v_i} {\cal H}_1 \\
- \p_{u_i} {\cal H}_1 \end{pmatrix}
\]
with the Hamiltonian
\[
{\cal H}_1= \mbox{tr}\, Y = \sum_{i=1}^{n}
e^{-v_i}\prod_{k\neq i}
\frac{u_i   -  u_k   +  1}{u_i -u_k} .
\]
This result was generalized to the dif\/ference
KP hierarchy (which is essentially equivalent to the
mKP hierarchy) in~\cite{Iliev}:
\begin{gather}\label{CM12}
 \begin{pmatrix}\p_{t_m}u_i \\ \p_{t_m} v_i \end{pmatrix}=
 \begin{pmatrix} \p_{v_i} {\cal H}_m \\
- \p_{u_i} {\cal H}_m \end{pmatrix},
\qquad   {\cal H}_m =\mbox{tr}\, Y^m.
\end{gather}
The ${\cal H}_m$'s are higher integrals of motion (Hamiltonians)
for the RS
model. They are known to be in involution~\cite{RS}. This agrees with the commutativity
of the mKP f\/lows.

For completeness, we give a derivation of~(\ref{CM12})
which is a version of the arguments from \mbox{\cite{Iliev, Shiota}}.
The main technical tool is equation (\ref{BA10}) which states that
\[
\sum_k \left (\frac{\p_{t_m}u_k}{u-u_k}-
\frac{\p_{t_m}u_k}{u-u_k+1}\right )=
\mbox{res}_{\infty}\left [ \left (c_0 +
\sum_i \frac{c_i}{u-u_i}\right )
\left ( c_{0}^{-1}+\frac{c_i^*}{u-u_i+1}\right ) z^{m-1}{\rm d}z \right ].
\]
Matching coef\/f\/icients in front of the poles, we get
\[
\p_{t_m}u_i=
-(\dot u_i)^{-1}\mbox{res}_{\infty}
\bigl (c_i c_i^*
z^m {\rm d}z \bigr ).
\]
Inserting here (\ref{cc}), we continue the chain of
equalities:
 \begin{gather*}
\p_{t_m}u_i  =
\mbox{res}_{\infty}
\left [\bigl ({\sf 1^t}(zI-Y)^{-1}\dot U\bigr )_i   (\dot u_i)^{-1}
\bigl ((zI-Y)^{-1}\dot U {\sf 1}\bigr )_i   z^m {\rm d}z \right ]\\
\hphantom{\p_{t_m}u_i}{}
 =
\mbox{res}_{\infty}
\left [\bigl ({\sf 1^t}(zI-Y)^{-1}\bigr )_i
\bigl ((zI-Y)^{-1}\dot U {\sf 1}\bigr )_i   z^m {\rm d}z \right ]\\
\hphantom{\p_{t_m}u_i}{}
 =
\mbox{res}_{\infty}
\left [{\sf 1^t}(zI-Y)^{-1}   E_{ii}
(zI-Y)^{-1}\dot U {\sf 1}   z^m {\rm d}z \right ]\\
\hphantom{\p_{t_m}u_i}{}
 =
\mbox{res}_{\infty}
\left [ \mbox{tr}\bigl ( (\dot U {\sf 1}\otimes {\sf 1^t})
(zI-Y)^{-1}   E_{ii}  (zI-Y)^{-1}\bigr ) z^m {\rm d}z \right ].
\end{gather*}
The next steps are to
use the commutation relation~(\ref{comm}) and
notice that
$ E_{ii}Y  =
\dot u_i \frac{\p Y}{\p \dot u_i}
$:
\begin{gather*}
\p_{t_m}u_i  =
\mbox{res}_{\infty}
\left [ \mbox{tr}\bigl ( (-Y+UY-YU)
(zI-Y)^{-1}   E_{ii}  (zI-Y)^{-1}\bigr ) z^m {\rm d}z \right ]\\
\hphantom{\p_{t_m}u_i}{}
 =   -  \mbox{res}_{\infty}
\left [ \mbox{tr}\left((zI-Y)^{-1}\frac{\p Y}{\p \log \dot u_i}
(zI-Y)^{-1}\right) z^m {\rm d}z \right ] \\
\hphantom{\p_{t_m}u_i=}{}
 + \mbox{res}_{\infty}
\left [ \mbox{tr}\bigl (E_{ii}(zI-Y)^{-1}(UY-YU)(zI-Y)^{-1}\bigr )
z^m {\rm d}z \right ].
\end{gather*}
The trace in the last term is
\begin{gather*}
\mbox{tr}\bigl (E_{ii}(zI-Y)^{-1}(UY-YU)(zI-Y)^{-1}\bigr )=
\mbox{tr}\bigl (E_{ii}\bigl [U,   (zI-Y)^{-1}\bigr ]\bigr )=
\bigl (\bigl [U,   (zI-Y)^{-1}\bigr ]\bigr )_{ii},
\end{gather*}
which is equal to zero because the matrix~$U$ is diagonal.
We are left with
\begin{gather*}
\p_{t_m}u_i  = -    \mbox{res}_{\infty}
\left [ \mbox{tr}\left ((zI-Y)^{-1}\frac{\p Y}{\p \log \dot u_i}
(zI-Y)^{-1}\right ) z^m {\rm d}z \right ]
\\
\hphantom{\p_{t_m}u_i}{} =
-   \mbox{res}_{\infty}
\left [ \frac{\p }{\p \log \dot u_i}\mbox{tr}\, \frac{1}{zI-Y}
z^m {\rm d}z \right ]
=
-   \frac{\p }{\p \log \dot u_i}\mbox{tr}\, Y^m
=  \p_{v_i} \mbox{tr} \, Y^m .
\end{gather*}
This proves the f\/irst equality in (\ref{CM12}).
Note that another form of the equation $\p_{t_m}u_i =
\p \, \mbox{tr} \, Y^m/\p v_i$ is
\begin{gather}\label{CM14a}
\p_{t_m}u_i =-m \, \mbox{tr}\big( E_{ii}Y^m \big)=
-m\big(Y^m\big)_{ii}.
\end{gather}

The proof of the second equality in (\ref{CM12}) is more
involved. Here we will closely follow \cite{Iliev}.
First, using the Lax equation
$\dot Y=[\tilde T, Y]$ and cyclicity of the trace,
we take the $t_1$-derivative of~(\ref{CM14a})
to get:
\[
\p_{t_m}\dot u_i =-m \, \mbox{tr}\big(Y^m [E_{ii}, \tilde T]\big)
\]
(recall that $\tilde T=T+Y$, see~(\ref{tildeT})).
With the help of this formula we can f\/ind $\p_{t_m}v_i$:
\begin{gather*}
\p_{t_m}v_i   =
-\dot u_{i}^{-1}\p_{t_m}\dot u_i +
\sum_{j=1}^{n}\sum_{l\neq i}\left(
\p_{u_j}\log \frac{u_i-u_l +1}{u_i-u_l}\right ) \p_{t_m}u_j
\\
\hphantom{\p_{t_m}v_i}{} =
m\dot u_{i}^{-1}\mbox{tr}\left (Y^m [E_{ii}, \tilde T]\right )-
m\sum_{j=1}^{n}\sum_{l\neq i}\left(
\p_{u_j}\log \frac{u_i-u_l +1}{u_i-u_l}\right )\mbox{tr}\bigl (Y^m
E_{jj}\bigr ) \\
\hphantom{\p_{t_m}v_i}{}
=  m\, \mbox{tr}\bigl (A^{(i)}Y^{m-1}\bigr ),
\end{gather*}
where
\[
A^{(i)}=\dot u_{i}^{-1}\bigl ( YE_{ii}\tilde T' -\tilde T'E_{ii}
Y\bigr ) -\sum_{j=1}^{n}\sum_{l\neq i}\left (
\p_{u_j}\log \frac{u_i-u_l +1}{u_i-u_l}\right)E_{jj}Y.
\]
Here $\tilde T'$ is the matrix $\tilde T$ (see~(\ref{Q}))
with zeros on the main diagonal, $\tilde T'_{ij}=\tilde T_{ij}-
\delta_{ij}\tilde T_{ii}$.
One can show that
\begin{gather}\label{CM16}
-A^{(i)}=\p_{u_i}Y +[C^{(i)}, Y]
\end{gather}
where $C^{(i)}$ is the matrix
\[
C^{(i)}=\sum_{l=1}^{n}\frac{E_{ll}}{u_{li}+1} -
\sum_{l\neq i}\frac{E_{ll}}{u_{li}}
\]
(here and below $u_{ij}\equiv u_i-u_j$).
From this it immediately follows that $\p_{t_m}v_i=\p_{u_i}
\mbox{tr} (Y^m  )$, which
is the second equality in~(\ref{CM12}).
The most direct way to prove~(\ref{CM16}) is to calculate matrix elements
of both sides.
For example, matrix elements of the matrix~$A^{(i)}$
are as follows:
\[
A^{(i)}_{jk}=-Y_{jk}\left (
\frac{1-\delta_{ij}}{u_{ik}-1}-\frac{1-\delta_{ik}}{u_{ik}}
-\frac{\delta_{ik}}{u_{ij}+1} +1 +\delta_{ij}
\sum_{l\neq i}\left (\frac{1}{u_{il}+1}-\frac{1}{u_{il}}\right )\right ).
\]

\subsection[Determinant formula for the master $T$-operator]{Determinant formula for the master $\boldsymbol{T}$-operator}

There is an explicit
determinant representation of the master $T$-operator.
Let $U_0 =U(0)$ be the diagonal matrix
$U_0=\mbox{diag}(u_1, u_2, \ldots , u_n)$,
where $u_i=u_i(0)$ and $Y_0$ be the
Lax matrix (\ref{CM6}) at ${\bf t}=0$, with
$\dot u_i (0) =-H_i$ (see~(\ref{CM2})).
Then
\begin{gather}\label{CM13}
T(u, {\bf t})=e^{\text{tr} \, \xi ({\bf t}, g)}
\det \left ( uI -U_0 +\sum_{k\geq 1}kt_k Y_0^{k}\right ).
\end{gather}
Substituting this into (\ref{BA1}), (\ref{BA2}) we f\/ind formulas for
the stationary BA functions:
\begin{gather}\label{BA1a}
\psi_u (z)=\det (zI - g) z^{u-N}
\frac{\det \bigl ((uI  -  U_0) (zI   -  Y_0)  -
  Y\bigr )}{\det (uI-U_0)\det (zI-Y_0)},
\\
\label{BA2a}
\psi^{*}_u (z)=\frac{z^{N-u}}{\det (zI - g)}
\frac{\det \bigl ((zI   -  Y_0)(uI  -  U_0)
  +  Y\bigr )}{\det (uI-U_0)\det (zI-Y_0)}.
\end{gather}

Let us show that these formulas are equivalent to the
stationary versions of (\ref{cc1}).
Using commutation relation (\ref{comm}), we have:
\begin{gather*}
  \det \bigl ( (uI-U_0)(zI-Y_0) - Y \bigr )
 =  \det \bigl ( (zI-Y_0)(uI-U_0)+[U_0,Y_0] - Y\bigr )
\\
\qquad{}= \det \Bigl ((zI-Y_0)(uI-U_0)+\dot U ({\sf 1}\otimes {\sf 1^t})\Bigr )
\\
\qquad{}=  \det (uI  -  U_0)   \det (zI  -  Y_0)  \det
\Bigl (I+ (uI-U_0)^{-1}(zI-Y_0)^{-1}\dot U ({\sf 1}\otimes {\sf 1^t})\Bigr )
\\
\qquad{}=  \det (uI  -  U_0)   \det (zI  -  Y_0)
\bigl(1+\mbox{tr}\bigl ((uI-U_0)^{-1}(zI-Y_0)^{-1}
\dot U ({\sf 1}\otimes {\sf 1^t})\bigr )
\\
\qquad{}=  \det (uI  -  U_0)   \det (zI  -  Y_0)
\bigl (1 + {\sf 1}^{\sf t}(uI-U_0)^{-1}(zI-Y_0)^{-1}\dot U
{\sf 1}\bigr )
\end{gather*}
and similarly for (\ref{BA2a}).
These formulas
show that (\ref{cc1}) and (\ref{BA1a}), (\ref{BA2a}) are
indeed equivalent.

Let us stress that determinant formulas
of the type (\ref{CM13}), (\ref{BA1a}) and (\ref{BA2a}) are not new
in the context of polynomial tau-functions
of classical integrable hierarchies
(see, e.g., \cite{Iliev, Shiota,Wilson}). The new observation is that
the master $T$-operator for quantum XXX spin chains has exactly this form.

\subsection{Spectrum of the
spin chain Hamiltonians from the classical RS model}

It follows from the above arguments that the eigenvalues of the
(non-local) spin chain Hamiltonians $H_i$, $i=1, \ldots , n$ (\ref{D2a}),
can be found in the framework of the classical
RS system with $n$ particles as follows. Consider the matrix
\begin{gather}\label{S1}
Y_0=  \begin{pmatrix}
H_1 & \dfrac{H_1}{u_2 - u_1 + 1} &
\dfrac{H_1}{u_3 - u_1 + 1} &
\ldots & \dfrac{H_1}{u_n - u_1 + 1}
\vspace{1mm}\\
 \dfrac{H_2}{u_1 - u_2 + 1} & H_2 &
 \dfrac{H_2}{u_3 - u_2 + 1} &
 \ldots & \dfrac{H_2}{u_n - u_2 + 1}
\vspace{1mm}\\
   \vdots & \vdots & \vdots & \ddots & \vdots
\vspace{1mm}\\
 \dfrac{H_n}{u_1 - u_n + 1} &
 \dfrac{H_n}{u_2 - u_n + 1}&
 \dfrac{H_n}{u_3 - u_n + 1} & \ldots & H_n
\end{pmatrix}.
\end{gather}
The spectrum of the $H_i$'s in the space ${\cal V}(\{ m_a\})$ is determined
by the conditions
\[
\mbox{tr}\, Y_0^j = \sum_{a=1}^{N} m_a w_a^j \qquad
\mbox{for all} \quad j\geq 1,
\]
i.e., given the initial coordinates $u_i$ and the action variables
${\cal H}_j =\mbox{tr}\, Y_0^j$ one has to f\/ind possible values
of the initial velocities $\dot u_i =-H_i$.
This
is equivalent to $n$ algebraic equations for $n$ quantities
$H_1, \ldots , H_n$.

In other words,
the eigenstates of the quantum Hamiltonians correspond to
the intersection points of two Lagrangian
manifolds in the phase space of the RS model.
One of them is the Lagrangian hyperplane
def\/ined by f\/ixing the $u_i$'s and
the other one is the Lagrangian sub\-mani\-fold obtained by
f\/ixing values
of the involutive integrals of
motion ${\cal H}_i$'s, with the latter being
determined by eigenvalues of the spin chain twist matrix.
This purely classical prescription appears to be
equivalent to the Bethe ansatz solution and
solves the spectral problem for the quantum spin chain.

\begin{Example}
Consider the vector ${\sf v}_a \in {\mathbb C}^N$ with
components $({\sf v}_a)_b =\delta_{ab}$.
Since $P_{ij} ({\sf v}_a)^{\otimes n} = ({\sf v}_a)^{\otimes n}$, the vector
$({\sf v}_a)^{\otimes n}$ is an eigenstate for the
Hamiltonians $H_i$ with the eigenvalues
\[
{w_a   \prod_{j=1, j\neq i}^n\frac{u_i-u_j+1}{u_i-u_j}}.
\]
It is also an eigenvector for the operators $M_b$ with
eigenvalues $m_b=n\delta_{ab}$. The matrix (\ref{S1}) in this case
is the $n\! \times \! n$
Jordan block with the only eigenvector ${\sf 1}$ with eigenvalue
$w_a$ and $\mbox{tr}\, Y_0^j =nw_a^j$.
\end{Example}

\subsection*{Acknowledgements}

The author thanks A.~Alexandrov,
A.~Gorsky, V.~Kazakov, S.~Khoroshkin, I.~Krichever, S.~Leu\-rent,
M.~Ol\-sha\-nets\-ky, A.~Orlov, T.~Ta\-ke\-be, Z.~Tsuboi, and A.~Zo\-tov
for discussions. Referees' remarks which helped to improve the
manuscript are gratefully acknowledged.
This work was supported in part
by RFBR grant 11-02-01220, by joint RFBR grants 12-02-91052-CNRS,
12-02-92108-JSPS and
by Ministry of Science and Education of Russian Federation
under contract 8207 and by
grant NSh-3349.2012.2 for support of
leading scientif\/ic schools.

\pdfbookmark[1]{References}{ref}
\LastPageEnding

\end{document}